\documentclass[pra,twocolumn, ,superscriptaddress,notitlepage,longbibliography]{revtex4-2}
\usepackage{amsmath}
\usepackage{amssymb}

\usepackage[unicode=true,colorlinks=true,citecolor=blue,urlcolor=blue]{hyperref}

\usepackage{bm}

\usepackage{graphicx}

\usepackage{multirow}

\usepackage[normalem]{ulem}

\usepackage{placeins}

\renewcommand {\phi}{{\varphi}}

\begin{document}

\title{
{Parametric resonant enhancement of motional entanglement under optimal control: an analytical study}
}

\author{Gad Horovitz}
\email{gad.horovitz@weizmann.ac.il}
\author{Alexander N. Poddubny}
\email{poddubny@weizmann.ac.il}
\affiliation{Department of Physics of Complex Systems, Weizmann Institute of Science, Rehovot 7610001, Israel}

\begin{abstract}
We study theoretically continuous-variable entanglement between the motional degrees of freedom of optically trapped massive particles, coupled via the Coulomb interaction, 
in the presence of a feedback control scheme.
We perform a detailed analysis of the parametric resonance induced by temporal modulation of the coupling strength, based on the system’s coupled nonlinear, nonhomogeneous dynamical equations.  
Our model accurately reproduces the numerical findings and provides closed-form expressions for the entanglement degree. We demonstrate that the stationary nonequilibrium entangled state is realized as a result of the competition between the parametric gain and the decoherence.
\end{abstract}
\maketitle 
\section{Introduction}\label{sec:intro}
Quantum entanglement continues to fascinate researchers even after almost a century since its discovery. While achieving entanglement between a few nanoscale quantum systems, such as individual spins \cite{petta2005coherent, Hensen2015Loophole}, atoms \cite{wilk2010entanglement, hofmann2012heralded}, or photons \cite{freedman1972experimental, aspect1982experimental, hong1987measurement} in a laboratory has now become routine, demonstrating entanglement between massive objects remains challenging.  
One of the promising platforms to this end is based on optically trapped dielectric nanoparticles. 
For now, cooling the trapped particles almost to the ground motional state \cite{Delic2020Cooling} and their Coulomb and dipole-dipole interaction between these particles has recently been demonstrated in the lab \cite{Rieser2022TunableInteraction}.  The next stage involves demonstrating continuous-variable entanglement between the motional degrees of freedom of the trapped particles. 
Coulombic entanglement of motional degrees of freedom of massive optically trapped nanoparticles is significant for various fields, including quantum metrology and even potentially fundamental physics tests \cite{Magrini2021RealTimeControl}. 
To achieve the entanglement, it is still necessary that the interparticle interaction overcomes the decoherence processes. A detailed theoretical study for state-of-the-art experimental parameters has recently been presented in Ref.~\cite{winkler2024stationary}. It turns out that the required strength of the coupling between the two particles $g$ is still unrealistically large. One of the ways to boost the entanglement is to employ the temporal modulation of the coupling~\cite{Galve2010HighTemperatureEntanglement,szorkovszky2014mechanical,lin2020entangling}. In a nutshell, this parametric modulation leads to a squeezing of the differential motion mode.  This approach  for the trapped particle system has been proposed and comprehensively numerically studied in  our previous work Ref.~\cite{poddubny2025nonequilibriumentanglementlevitatedmasses}. The essential distinction of Ref.~\cite{poddubny2025nonequilibriumentanglementlevitatedmasses} from the earlier studies of parametric  entanglement generation~\cite{Galve2010HighTemperatureEntanglement}
has been taking into account continuous interferometric position measurement of the particles. In particular, the proposed protocol employed feedback control to stabilize the effective temperature of the motion, optimizing entanglement likelihood through a specifically chosen cost function \cite{Edwards_2005, hofer2015entanglement}. The results of Ref.~\cite{poddubny2025nonequilibriumentanglementlevitatedmasses} suggested that for the attractive interaction between the two particles, the parametric modulation with the depth of 20\% and a realistic detection efficiency of 50\% leads to entanglement at the interaction strength \(g_0=0.1\Omega_0\)~--~more than an order of magnitude improvement over existing stationary protocols~\cite{Rudolph2022ForceGradientSensing,winkler2024stationary}. At the same time, in the repulsive regime, only stroboscopic unconditional entanglement and stationary conditional entanglement have been attained. However, the analysis in Ref.~\cite{poddubny2025nonequilibriumentanglementlevitatedmasses} has been mostly numerical.  
The fundamental theoretical problem of continuous-variable entanglement generation in the parametrically driven systems in the presence of feedback control has still not been fully analyzed to the best of our knowledge.  
\begin{figure}[b]
\centering\includegraphics[width=0.49\textwidth]{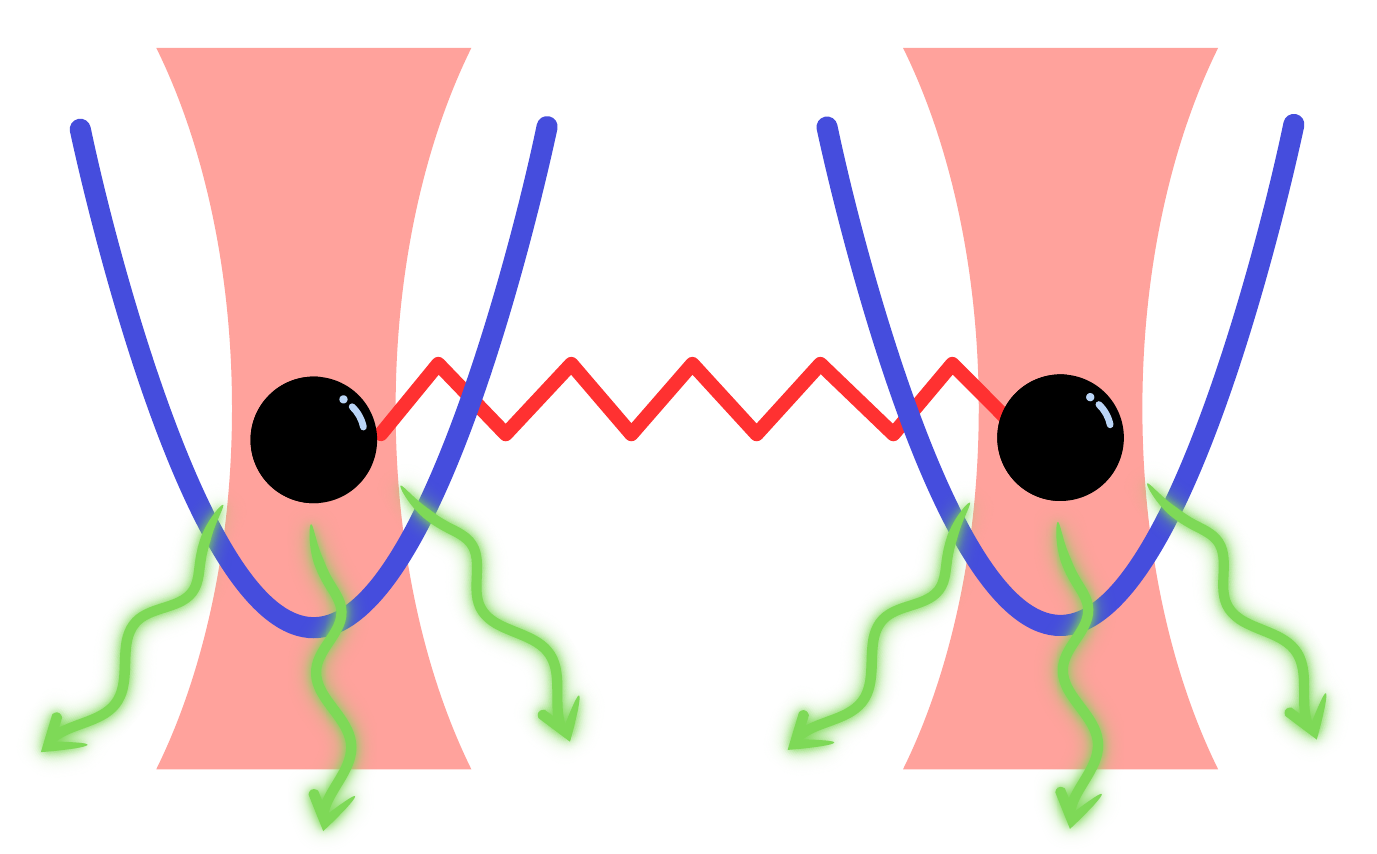}
\caption{Schematic illustration of two interacting optically trapped particles. }
\label{fig1}
\end{figure}

In this work, we put forward a detailed theoretical model to characterize and predict the strength and behavior of the entanglement in the presence of feedback control and parametric modulation. Our model is based on the investigation of the parametric resonance in the system’s coupled nonlinear, nonhomogeneous dynamical equations for the covariance matrix and accurately semianalytically reproduces the numerical findings. We obtain seminanalytical results for both conditional and unconditional negativity, quantifying the entanglement, as well as explicit analytical results for the conditional negativity in the resonance driving case. This allows us to quantitatively investigate the negativity dependence on the decay rates, the parametric modulation strength, and the interferometric measurement efficiency.

The rest of this work is organized as follows. We start in Sec.~\ref{sec:system} by outlining our general theoretical model used for numerical calculations. The notation and equations closely follow Ref.~\cite{poddubny2025nonequilibriumentanglementlevitatedmasses} and Ref.~\cite{winkler2024stationary}, as well as earlier Ref.~\cite{hofer2015entanglement}. 
Next, in Sec.~\ref{sec:analytics} we present the analytical model for the covariance matrix and entanglement, based on the analysis of the parametric resonance in the framework of a simple  Mathieu equation for the differential mode of the two particles. We also compare the analytical results with numerical simulations. The results are summarized in Sec.~\ref{sec:summary}.

\section{Theoretical model}\label{sec:system}

As in Refs.~\cite{winkler2024stationary,poddubny2025nonequilibriumentanglementlevitatedmasses}, the considered system is comprised of two charged, optically trapped, identical particles with mass $m$, separated by a mean distance $R$ (see Fig.~\ref{fig1}). Each of the particles is subject to a harmonic potential with frequency $\Omega_{0}$, induced by the trap along the optical axis and orthogonal to $R$, with position coordinates $x_{1,2}=\sqrt{\hbar/m\Omega_{0}}X_{1,2} \equiv x_{\text{zpf}}X_{12}$, such that the total distance between the particle is $r = \sqrt{R^{2} + x_{\text{zpf}}^{2} (X_{1} - X_{2})^{2}}$. 
For simplicity,  we consider only the one-dimensional problem, where the particles move along $x$. The particles are electrically charged and experience a Coulombic interaction characterized by the Hamiltonian of the form $H_{\text{int}}=k/r$. Assuming small displacements under the harmonic potential, the Hamiltonian can be approximated as   $H_{\text{int}} \approx\hbar g(X_{1} - X_{2})^{2}$ with $g=-k / (2R^{3} m \Omega_{0})$ serving as the interaction's coupling rate. In this work, we let the interparticle distance $R$ be modulated such that the interaction rate $g$ will have an effective form
\begin{equation}
    g(t) = g_{0} + 2g_{1} \cos{\Omega_{c} t}\:,
    \label{eq: g(t))}
\end{equation}
with $g_{1},\Omega_{c}$ being the amplitude and frequency of the modulation, respectively. Lastly, a feedback force of the form $-\hbar u_{1,2}(t)X_{1,2}$ is applied to each particle. The total Hamiltonian then takes the form 
\begin{equation}
\begin{split}
    H &= \frac{\hbar \Omega_{0}}{2} \sum_{i=1,2} \left[\left( P_{i}^{2} + X_{i}^{2} \right) - \hbar u_{i} \left(t\right) X_{i} \right]
    \\& + \hbar g\left(t\right) \left( X_{1} - X_{2}\right)^{2},
\end{split}
\label{eq:Hamiltonian_no_units}
\end{equation}
with $p_{1,2} =\sqrt{\hbar m \Omega_{0}} P_{1,2} \equiv p_{\text{zpf}} P_{1,2}$. The dynamics of the system can be decoupled into two separate subsystems by a transformation into  common and differential modes
\begin{equation}
    x_{\pm} = \frac{x_1 \pm x_2}{\sqrt{2}}, \qquad  p_{\pm} = \frac{p_1 \pm p_2}{\sqrt{2}} 
    \label{eq: modes transformation},
\end{equation}
with the eigenfrequencies 
\begin{equation}
\Omega_{+} = \Omega_{0}, \qquad \Omega_{-}=\sqrt{\Omega_{0}^{2}+4\Omega_{0} g_{0}}\:. 
\label{eq: eigenmodes}
\end{equation}

Apart from the coherent evolution described above, the system experiences two dissipative processes in the form of collisions with residual gas molecules and photon scattering. The first process is characterized by the decoherence rate $\Gamma_{\text{th}}$, and the second by the rate $\Gamma_{\text{ba}}$. The thermal decoherence rate is related to the mechanical damping rate, $\gamma$, and to the mean phonon occupation number  $\Gamma_{\text{th}} = \gamma \bar{n} $. At room temperature, the occupation number can be approximated as $\bar{n} \approx k_{B} T / \hbar \Omega_{0}$.

The position and momentum of the system are continuously measured by homodyne detection of the scattered photons. The measurement gives rise to the conditioned state $\rho_{c}$, whose dynamics are given by the master equation \cite{Jacobs_2014, winkler2024stationary} 
\begin{multline}
     d\rho_{c} = -\frac{i}{\hbar}\left[H,\rho_{c}\right]dt-\sum_{k=1,2}\mathcal{D}_{\text{th}}^{k}\left[\rho_{c}\right]dt-\sum_{k=1,2}\mathcal{D}_{\text{ba}}^{k}\left[\rho_{c}\right]dt
    \\+\sum_{k=1,2}\mathcal{H}^{k}\left[\rho_{c}\right]dW_{k}\:,
\label{eq: stochastic_ME}
\end{multline}
where $\mathcal{D}_{\text{th}}$ describes the collision dissipation under the Brownian motion Caldeira-Leggett model \cite{CALDEIRA1983587}, $\mathcal{D}_{\text{ba}}$ describes photon recoil dissipation \cite{Rudolph2001Theory}, and $\mathcal{H}$ is the Belavkin term \cite{Edwards_2005}, describing the stochastic measurement process alongside the Wiener increment, which satisfies $dW_{k}$,  $\mathbb{E}[dW_{k}] = 0$, $\mathbb{E}[dW_{k},dW_{l}] = \delta_{k,l} dt/2$. The complete forms of the three processes are given by
\begin{gather}
\mathcal{D}_{\text{th}}^{k}\left[\rho_{c}\right]=\frac{i\gamma}{2}\left[x_{k},\left\{ p_{k},\rho_{c}\right\} \right]+\Gamma_{\text{th}}\left[x_{k},\left[x_{k},\rho_{c}\right]\right]\:,
\label{eq: Lindblad_th}\\
\mathcal{D}_{\text{ba}}^{k}\left[\rho_{c}\right]=\Gamma_{\text{ba}}\left[x_{k},\left[x_{k},\rho_{c}\right]\right]\:,
\label{eq: Lindblad_ba}\\
\mathcal{H}^{k}\left[\rho_{c}\right]=\sqrt{2\Gamma_{m}}\left\{ x_{k}-\left\langle x_{k}\right\rangle _{c},\rho_{c}\right\},
\label{eq: Belavkin}
\end{gather}
where $\Gamma_{m}$ is the detection rate, and is related to $\Gamma_{ba}$ by the measurement efficiency $\eta = \Gamma_{\text{m}} / \Gamma_{\text{th}}$. 

The quadratic form of Eq.~\eqref{eq: stochastic_ME} ensures that an initial Gaussian state remains Gaussian throughout its evolution. Consequently, the state can be fully characterized by its first and second quadrature moments. Denoting the quadrature moment as $\boldsymbol{X} = \left( X_{1}, P_{1}, X_{2}, P_{2} \right)^{T}$, the conditional first moment is formally given by  
\begin{equation}
    \boldsymbol{X}_{c}  = \langle \boldsymbol{X} \rangle_{c} = \text{Tr} \left[ \boldsymbol{X} \rho_{c} \right].
    \label{eq: first moment}
\end{equation}
The conditional second moment is represented by the positive definite \cite{Serafini2017quantum}  covariance matrix, $\Sigma^{c}$, defined as  
 \begin{equation}
    \Sigma^{c} = \text{Re} \langle \left( \boldsymbol{X} - \boldsymbol{X}_{c} \right) \left( \boldsymbol{X} - \boldsymbol{X}_{c} \right)^T \rangle_{c}.
\label{eq:covariance}
\end{equation}
The dynamics of the first and second moments are dictated by \eqref{eq: stochastic_ME}, and are given by 
\begin{align}
\begin{split}
    d\boldsymbol{X}_{c}\left(t\right) &= \boldsymbol{A}\left(t\right)\boldsymbol{X}_{c}\left(t\right)dt+\boldsymbol{B}\boldsymbol{u}\left(t\right)dt \\
    &+ \Sigma^{c}\left(t\right)\boldsymbol{C}^{T}\boldsymbol{W}^{-1}d\boldsymbol{w}(t)\:,
\end{split}\label{eq: mean_EOM}\\
\begin{split}    d\Sigma^{c}\left(t\right)&=\boldsymbol{A}\left(t\right)\Sigma^{c}\left(t\right)dt+\Sigma^{c}\left(t\right)\boldsymbol{A}^{T}\left(t\right)dt+\boldsymbol{V}dt \\
    &-\Sigma^{c}\boldsymbol{C}^{T}\left(t\right)\boldsymbol{W}^{-1}\boldsymbol{C}\Sigma^{c}\left(t\right)dt\:,
\end{split}\label{eq: covariance_EOM}
\end{align}
where $\boldsymbol{A}(t)$ is the drift matrix, $\boldsymbol{B}$ the control matrix, $\boldsymbol{u}(t)$ the input force vector, $\boldsymbol{C}$ the measurement matrix, $\boldsymbol{V}$ the system noise correlation matrix, $\boldsymbol{W}$ the measurement noise correlation matrix, and $d\boldsymbol{w}(t)$ a stochastic vector. In the normal mode basis, the explicit decoupled forms of the elements of Eqs.~\eqref{eq: mean_EOM} and \eqref{eq: covariance_EOM} are
\begin{gather}
    \Sigma_{c}=\begin{pmatrix}\boldsymbol{\Sigma}_{c}^{+} & \boldsymbol{0}_{2\times2}\\
    \boldsymbol{0}_{2\times2} & \boldsymbol{\Sigma}_{c}^{-}
    \end{pmatrix}, \qquad \boldsymbol{\Sigma}_{c}^{\pm}=\begin{pmatrix}\Sigma_{c,11}^{\pm} & \Sigma_{c,12}^{\pm}\\
    \Sigma_{c,12}^{\pm} & \Sigma_{c,22}^{\pm}
    \end{pmatrix}\:,
\label{eq: covariane matrix}\\
    \boldsymbol{A}\left(t\right)=\begin{pmatrix}0 & \Omega_{0} & 0 & 0\\
    -\Omega_{0} & -\gamma & 0 & 0\\
    0 & 0 & 0 & \Omega_{0}\\
    0 & 0 & -\Omega_{0}-4g\left(t\right) & -\gamma
\end{pmatrix}\:,
\label{eq: Drift matrix}\\
    \boldsymbol{B}=\begin{pmatrix}0 & 0\\
    1 & 0\\
    0 & 0\\
    1 & 0 
\end{pmatrix}\:, \qquad \boldsymbol{u}\left(t\right)=\begin{pmatrix}u_{+}\left(t\right)\\
    u_{-}\left(t\right) 
    \end{pmatrix}\:, 
\label{eq: Control matrix}\\
    \boldsymbol{C}=\sqrt{\Gamma_{m}}\begin{pmatrix}1 & 0 & 0 & 0\\
    0 & 0 & 1 & 0
    \end{pmatrix}\:, \qquad  d\boldsymbol{w}(t) = \begin{pmatrix}
        dW_{+}(t)\\ dW_{-}(t)
    \end{pmatrix}
\label{eq: Stochastic vector & measurement matri}\\
    \boldsymbol{W}=\frac{1}{2}\begin{pmatrix}
        1&0\\0&1
    \end{pmatrix}\:, \qquad 
\label{eq: Measurement noise}
\end{gather}
where $dW_{\pm}(t)$ describes a zero mean Wiener increment with $\mathbb{E}\left[ dW_{i} dW_{j} \right]  = \delta_{ij} dt/2$. We also note that the form of~Eq.~\eqref{eq: Control matrix} allows for individual control of each particle.

Insofar, we had described the \textit{conditional} dynamics of the system, whose first moment is intrinsically non-deterministic, and will explore a different phase space trajectory at each realization.  We are interested, however, in the deterministic and measurable unconditional dynamics of the system, whose state is defined as the ensemble average over all conditional states, i.e. $\rho = \mathbb{E}[\rho_{c}]$. The unconditional covariance matrix associated with $\rho$ can be decoupled into the conditional covariance matrix, $\Sigma$, and an excess noise matrix, $\Xi$ \cite{hofer2015entanglement}: 
\begin{equation}
    \Sigma=\Sigma_{c}+\Xi.
\label{eq: unconditional covariance}
\end{equation}
From the above equation, it follows that the conditional covariance, $\Sigma_{c}$, acts as a  lower bound to the total, unconditional covariance. The force-feedback term can be employed to steer the unconditional state toward a more desirable configuration by minimizing the excess-noise contribution in Eq.~\eqref{eq: unconditional covariance} using a linear quadratic regulator (LQR)~\cite{winkler2024stationary}
\begin{equation}
    \mathcal{J} = \int_{t_{0}}^{T} \mathbb{E} \left[ \boldsymbol{X}^T_{c} \boldsymbol{P} \boldsymbol{X}_{c} + \boldsymbol{u}^T q \boldsymbol{u}  \right] dt,
\label{eq: LQR}
\end{equation}
where $\boldsymbol{P}$ is a cost matrix, and $q$ denotes the control effort. Following \cite{winkler2024stationary}, we choose the EPR cost matrix 
\begin{equation}
    \boldsymbol{P}=\Omega_{0}\begin{pmatrix}\boldsymbol{P}\left(\theta\right) & \boldsymbol{0}_{2\times2}\\
    \boldsymbol{0}_{2\times2} & \boldsymbol{P}\left(\theta+\pi\right)
\end{pmatrix},
\label{eq: cost matrix 4x4}
\end{equation}
with 
\begin{equation}
    \boldsymbol{P} = \left(\theta\right)\begin{pmatrix}1+\cos\left(\theta\right) & \sin\left(\theta\right)\\
    \sin\left(\theta\right) & 1-\cos\left(\theta\right)
\end{pmatrix}.
\label{eq: cost matrix 2x2}
\end{equation}
From control theory \cite{hofer2015entanglement}, the minimizer input force of Eq.~\eqref{eq: LQR} is given by
\begin{equation}
    \boldsymbol{u} (t) = -\frac{ \boldsymbol{B}^T \Omega(t)}{q} \boldsymbol{X}_{c}(t) \equiv -\boldsymbol{K}(t) \boldsymbol{X}_{c}(t),
\label{eq: u minimizer}
\end{equation}
where the dynamics of $\Omega(t)$ are given by 
\begin{equation}
\begin{split}
    -d\Omega\left(t\right) &=\boldsymbol{A}\left(t\right)\Omega\left(t\right)dt+\Omega\left(t\right)\boldsymbol{A}^{T}\left(t\right)dt+\boldsymbol{P}\left(t\right)dt\\&
    -\Omega\left(t\right) \frac{\boldsymbol{B}\boldsymbol{B}^{T}}{q} \Omega\left(t\right)dt.
\end{split}
\label{eq: gain matrix eom}
\end{equation}
Finally, dynamics of the excess noise, $\Xi$, depends on both $\Sigma_{c}$ and $\Omega$ as \cite{hofer2015entanglement}
\begin{equation}
\begin{split}
    d\Xi\left(t\right) &=\left[\boldsymbol{A}\left(t\right)-\boldsymbol{B}\boldsymbol{K}\left(t\right)\right]\Xi\left(t\right)dt \\&+\Xi\left(t\right)\left[\boldsymbol{A}\left(t\right)-\boldsymbol{B}\boldsymbol{K}\left(t\right)\right]^{T}dt \\&+\Sigma^{c}\left(t\right)\boldsymbol{C}^{T}\boldsymbol{W}^{-1}\boldsymbol{C}\Sigma^{c}\left(t\right)dt.
\end{split}
\label{eq: excess noise EOM}
\end{equation}
We note that just like \ in the normal mode basis, $\Omega$ and $\Xi$ decouple into $2\times2$ subblocks of $\Omega^{\pm}$ and $\Xi^{\pm}$, such that the total unconditional covariance matrix is also decoupled into $\Sigma^{\pm}$ as well.
The entanglement of the system is quantified by the logarithmic negativity, defined as $\mathcal{E_{N}} = -\ln(2\tilde{\nu})$ \cite{Adesso_2007, Werner_2002}, where $\tilde{\nu}$ is the minimal symplectic eigenvalue of the partially transposed covariance matrix. The logarithmic negativity is defined for both the conditional and unconditional states, and indicates entanglement for positive values. In the normal modes basis, $\tilde{\nu}$ can be expressed as \cite{winkler2024stationary}
\begin{equation}
    \tilde{\nu}= \frac{1}{\sqrt{2}} \sqrt{\sigma- \sqrt{\sigma^{2} - 4 \left|\Sigma ^{+}\right| \left|\Sigma ^{-}\right|}}, 
\label{eq: symplectic eigenvalue}
\end{equation}
where 
\begin{equation}
    \sigma = \Sigma_{11}^{+}\Sigma_{22}^{-}+\Sigma_{22}^{+}\Sigma_{11}^{-}-2\Sigma_{12}^{+}\Sigma_{12}^{-}.
\label{eq: sigma negativity}    
\end{equation}

\section{Analytical theory}\label{sec:analytics}
In this section, we present approximate analytical solutions to the general covariance-matrix equations outlined above.
In Sec.~\ref{sec:system}, we showed that the system can be decoupled into common and differential subsystems, with the time dependence residing solely in the differential mode. The covariance of the common mode therefore reaches a steady state solution, reducing Eq.~\eqref{eq: covariance_EOM} to an analytically solvable algebraic equation, which was shown in Ref.~\cite{winkler2024stationary} to be
\begin{gather}
\begin{split}
         &\Sigma^{+}_{c,11} = -\frac{\gamma}{2 \eta\Gamma_{\text{ba}}} 
    \\& + \frac{1}{2 \eta\Gamma_{\text{ba}}} 
    \sqrt{ \gamma^{2} + 2 \Omega_{0} 
    \left( \sqrt{ 2 \eta\Gamma_{\text{ba}} \Gamma_{\text{tot}} +  \Omega_{0}^{2}  } -  \Omega_{0} \right)}\:,
    \end{split}
\label{eq: sigma_xx} \\
    \Sigma^{+}_{c,12} = \frac{\eta\Gamma_{\text{ba}}}{ \Omega_{0}}  \left( \Sigma^{+}_{c, 11} \right)^2\:,
\label{eq: sigma_xp}\\
\Sigma^{+}_{c,22} =
\left(
\frac{ \sqrt{ 2 \Gamma_m \Gamma_{\text{tot}} +  \Omega_{0}^{2} } }{\Omega_{0}}
- \frac{\eta\Gamma_{\text{ba}} \gamma}{ \Omega_{0}^{2}} \Sigma^{+}_{x,11}
\right) \Sigma^{+}_{x,11}.
\label{eq: sigma_pp}
\end{gather}

The differential mode poses a separate challenge. Its conditional second moment dynamics are nonlinear, nonhomogeneous, and explicitly time dependent, as dictated by Eq.~\eqref{eq: covariance_EOM}, and therefore cannot be solved analytically without approximations. The first approximation assumes weak nonlinearity and strong resonance, such that the system’s behavior is primarily governed by the linear part of Eq.~\eqref{eq: covariance_EOM}. This requires $\Omega_{0} \gg \Gamma_m$ and $\Omega_c \approx 2\Omega_{-}$, consistent with typical experimental parameters reported in Ref.~\cite{winkler2024stationary}. The second approximation, also supported in Ref.~\cite{winkler2024stationary}, assumes $\gamma \to 0$, i.e., $\gamma$ is much smaller than all other system parameters. Under these approximations, the linear part of Eq.~\eqref{eq: covariance_EOM} can be written as
\begin{equation}
    \frac{d\Sigma_{l}\left(t\right)}{dt}=\boldsymbol{A}_{l}\left(t\right)\Sigma_{l}\left(t\right)+\Sigma_{l}\left(t\right)\boldsymbol{A}_{l}^{T}\left(t\right),
\label{eq: linear EOM}
\end{equation} 
where $\Sigma_{l}$ denotes the solution of the linear equation~\eqref{eq: linear EOM}, and $\boldsymbol{A}_{l}(t)$ is the drift matrix of the differential mode in the absence of damping.
\begin{equation}
    \boldsymbol{A}_{l}\left(t\right)=\begin{pmatrix}0 & \Omega_{0}\\
    -\Omega_{0}-4g\left(t\right) & 0
\end{pmatrix}.
\label{eq: linear A(t)}
\end{equation}
Equation~\eqref{eq: linear EOM} can be simplified by letting $\Sigma_{l} = \boldsymbol{X}_{l} \boldsymbol{X}_{l}^{T}$, where $\boldsymbol{X}_{l} = \left( X_{l}, P_{l} \right)^{T}$ is an auxiliary two dimensional vector, whose equation of motion is given by 
\begin{equation}
    \frac{d \boldsymbol{X}_l}{dt} = \boldsymbol{A}(t)\boldsymbol{X}_l
\label{eq: X_l EOM}.
\end{equation}
Equation~\eqref{eq: linear EOM} can be recovered by taking the time derivative of $\Sigma_{l} = \boldsymbol{X}{l}\boldsymbol{X}{l}^{T}$. Using the explicit form of $\boldsymbol{A}{l}(t)$ given in Eq.~\eqref{eq: linear A(t)}, we find that the vector $\boldsymbol{X}{l}$ evolves as
\begin{equation}
    \frac{d^{2}{X}_{l}(t)}{dt^{2}} =-\Omega_{-}^{2}\left(1+\frac{8\Omega_{0}g_{1}}{\Omega_{-}^{2}}\cos\left(\Omega_{c}t\right)\right) X_{l}(t),
\label{eq: Mathieu_special}
\end{equation}
and $P_{l}$  as 
\begin{equation}
    P_{l}(t) = \frac{1}{\Omega_{0}} \frac{d X(t)}{dt}\:.
\label{eq: P_l EOM}
\end{equation}
Equation~\eqref{eq: Mathieu_special} corresponds to the well-known Mathieu equation and can be rewritten as
\begin{equation}
    \ddot{x}(t)=-\Omega_{-}^{2}\left\{1+h\cos\left[\left(2 +\epsilon\right) \Omega_{-} t\right])\right\} x(t),
\label{eq: Mathieu _general}
\end{equation}
with the parametric modulation strength being characterized by the dimensionless parameter $h=8\Omega_{0} g_{1} / \Omega_{-}^{2}$, and $\Omega_{c} = \left(2 +\epsilon\right) \Omega_{-} $. According to the general theory of the Mathieu equation \cite{Kovacic2018Mathieu}, for weak modulation $( h \ll 1)$, the system will exhibits resonance for the frequency range $-\frac{1}{2}h<\epsilon<\frac{1}{2}h$. Within the resonance range, the system's dynamics will approximately split into two $\Omega_{c}/2$-periodic modes multiplied by an exponential envelope, one diverging and the other decaying as (see Appendix \ref{app: modes derivation}) 
\begin{equation}
\begin{split}
        & e^{\frac{\mu t}{2}} \boldsymbol{X}_{\text{div}} (t)= e^{\frac{\mu t}{2}} \begin{pmatrix}\cos\left(\frac{\Omega_{c}}{2}t+\phi\right)\\-\frac{\Omega_{c}}{2 \Omega_{0}}\sin\left(\frac{\Omega_{c}}{2}t+\phi\right)\end{pmatrix}\:,
        \\& e^{-\frac{\mu t}{2}} \boldsymbol{X}_{\text{dec}} (t) = e^{-\frac{\mu t}{2}}\begin{pmatrix}\cos\left(\frac{\Omega_{c}}{2}t-\phi\right)\\ -\frac{\Omega_{c}}{2 \Omega_{0}}\sin\left(\frac{\Omega_{c}}{2}t-\phi\right)\end{pmatrix},
\end{split}
\label{eq: Mathieu solutions}
\end{equation}
with the Mathieu exponent $\mu=\Omega_{-}\sqrt{\frac{h^{2}}{4}-\epsilon^{2}}$ and the relative phase $\phi$, defined as $\tan \phi = \sqrt{\frac{h - 2\epsilon}{h + 2\epsilon}}$. 

\begin{figure}
\centering\includegraphics[width=0.49\textwidth]{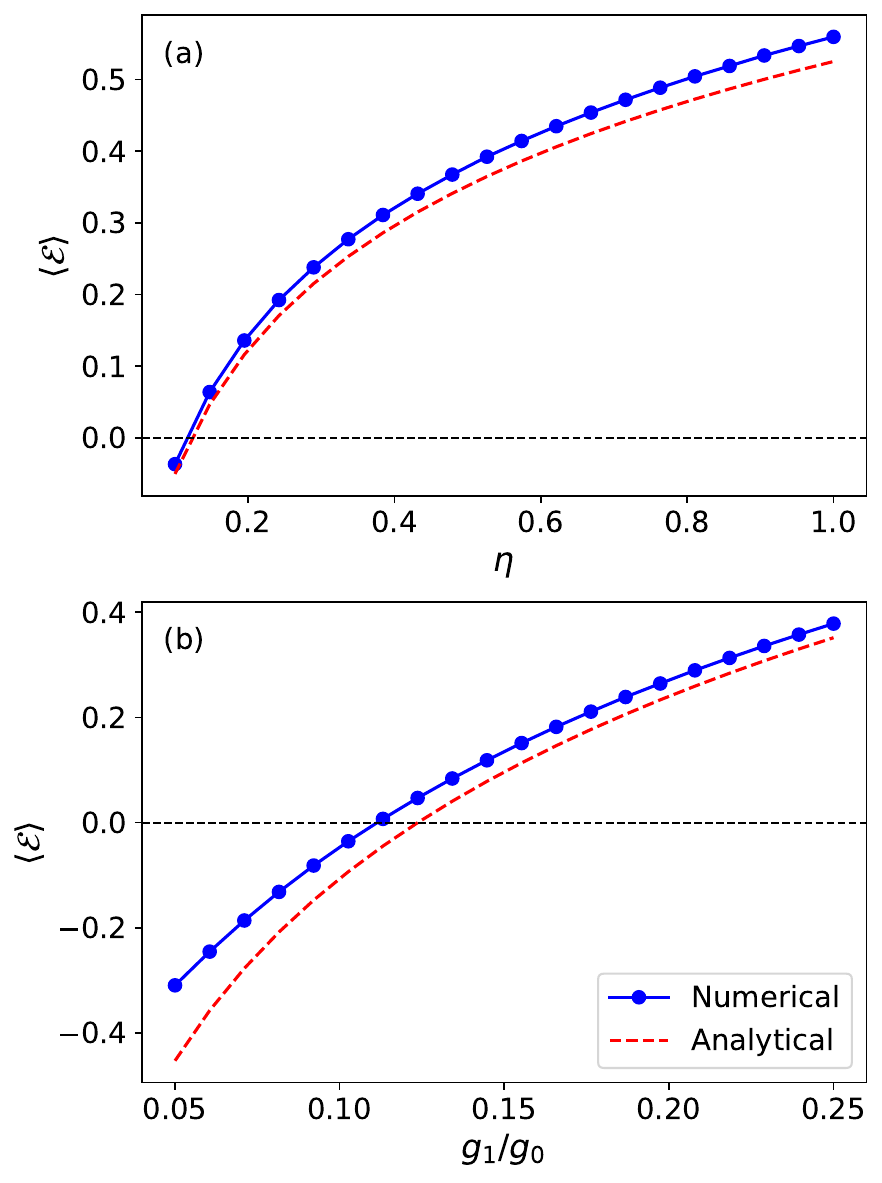}
\caption{Attractive interaction logarithmic negativity for the modulation frequency  $\Omega_{c}=2 \Omega_{-}$ as a function of the efficiency $\eta$ (a) and the modulation depth $g_{1}$ (b) averaged over one period. Solid line with markers denotes the numerical results, and the dashed line describes the value of the analytical expression of the negativity Eq.~\eqref{eq: approximate logarithmic negativity}. The black dashed line indicates the entanglement threshold. System parameters are: $\Omega_{c} = 2\Omega_{-}$, $\Omega_0/2\pi = 29.4\,\text{kHz}$, $g_{0}=0
.2\Omega_{0}$, $\Gamma_{\rm ba} = 1.3\,\text{kHz}$, $\Gamma_{\rm th} = 66.2\,\text{Hz}$, $\gamma = 0.31\,\mu\text{Hz}$, $Q = 1.08\,\mu\text{Hz}$, and $\theta_{\rm EPR} = \pi$.}
    \label{fig: g1-eta logneg}
\end{figure}

The exact solution of the conditional covariance matrix does not decay or diverge, as the nonlinear measurement term of Eq.~\eqref{eq: covariance_EOM} prevents it from doing so. However, since the nonlinearity is weak, $\Gamma_{m}\ll \Omega_{0}$,  we assume that the solutions of the Mathieu equation still characterize the dynamics of the system, with their exponential envelope suppressed by the nonlinearity, thereby yielding a conditional covariance matrix of the form 
\begin{equation}
    \Sigma^{-}_{c}\left(t\right)\approx\alpha_{\text{div}}\Sigma_{\text{div}}\left(t\right)+\alpha_{\text{dec}}\Sigma_{dec}\left(t\right),
\label{eq: Conditional approximation}
\end{equation}
with $\Sigma_{\text{div}}(t)=\boldsymbol{X}_{\text{div}}(t)\boldsymbol{X}_{\text{div}}^{T}(t)$, $\Sigma_{\text{dec}}(t)=\boldsymbol{X}_{\text{dec}}(t)\boldsymbol{X}_{\text{dec}}^{T}(t)$   being  the diverging and decaying modes of Eq.~\eqref{eq: X_l EOM}, albeit without their exponential term. Here,  $\alpha_{\text{div}}, \alpha_{\text{dec}}$ are the amplitudes of the two modes, which are to be determined. We expect the diverging mode to be the dominant one, i.e.  $\alpha_{\text{div}} \gg \alpha_{\text{dec}}$. Consequently, we assume $\alpha_{\text{div}}$ to be independent from $\alpha_{\text{dec}}$, though not the other way around. The amplitude of the diverging mode can be found by considering the structure of Eq.~\eqref{eq: covariance_EOM}. The first and linear term of the right-hand side, whose strength is characterized by the Mathieu exponent $\mu$, will compete with the restoring-like effect of the nonlinear term of the right-hand side, with the strength of  $\Gamma_{\text{m}}=\eta \Gamma_{\text{ba}}$. At the periodic steady state, we  assume the amplitude to be the ratio between the two
\begin{equation}
    \alpha_{\text{div}}= \frac{\mu }{\eta \Gamma_{\text{ba}}}\:.
\label{eq: amp_div}
\end{equation}
To determine the amplitude of the decaying mode $\alpha_{\text{dec}}$, we rely on the fact that the determinant of each mode, $\Sigma^{\pm}_{c}(t)$, will always converge to a steady state value. This statement can be proved directly by taking the time derivative of the determinant, $\left| \Sigma^{\pm}_{c}(t) \right|$,
\begin{equation}
\frac{d}{dt}\left|\boldsymbol{\Sigma}_{c}^{\pm}\left(t\right)\right|=\left[\left(\Gamma_{\text{th}}+\Gamma_{\text{ba}}\right)-2\eta\Gamma_{\text{ba}}\left|\Sigma\left(t\right)\right|\right]\Sigma_{11}^{\pm},
\label{eq: determinnat derivative}
\end{equation}
and noting that since $\Sigma_{11}^{\pm} >0$, the sign of the derivative will not be affected by it, and thus the determinant will flow toward the stable fixed point
\begin{equation}
\left|\boldsymbol{\Sigma}_{c}^{\pm}\right|=\frac{\Gamma_{\text{th}}+\Gamma_{\text{ba}}}{2\eta\Gamma_{\text{ba}}}\:.
\label{eq: ellipse area}
\end{equation}
We also calculate the determinant directly using~Eq.~\eqref{eq: Mathieu solutions} and~Eq.~\eqref{eq: Conditional approximation}, obtaining
\begin{equation}    
    \left|\boldsymbol{\Sigma}_{c}^{\pm}\right|=\left(\frac{\Omega_{c}}{2\Omega_{0}}\right)^{2}\sin^{2}\left(2\phi\right)\alpha_{\text{div}}\alpha_{\text{dec}}.
\label{eq: determinnat sum}
\end{equation}
By equating~Eq.~\eqref{eq: ellipse area} to~Eq.~\eqref{eq: determinnat sum}, and substituting~Eq.~\eqref{eq: amp_div}, we can write $\alpha_{\text{dec}}$ as
\begin{equation}
\alpha_{\text{dec}}=2\left(\frac{\Omega_{0}}{\Omega_{c}}\right)^{2}\frac{ \Gamma_{\text{ba}} + \Gamma_{\text{th}} }{\eta\mu\sin^{2}\left(2\phi\right)}.
\label{eq: amp_dec}
\end{equation} 
While in principle our theory allows for a complete analytical expression of $\mathcal{E_{N}}$ at any modulation frequency $\Omega_{c}$ within the resonance range, the resulting formula is cumbersome and offers limited additional insight. However, under exact resonance modulation of $\Omega_{c} = 2\Omega_{-}$, the logarithmic negativity can be presented in a more useful and compact way. This can be done by noting that under the approximations introduced at the beginning of Sec.~\ref{sec:analytics}, the conditional covariance of the common mode can be approximated from equations~\eqref{eq: sigma_xx},~\eqref{eq: sigma_xp} and~\eqref{eq: sigma_pp} to the simpler form 
\begin{equation}
\begin{split}
    &\Sigma^{+}_{c,11} \approx \Sigma^{+}_{c,22} \approx \sqrt{ \frac{\Gamma_{\text{tot}}}{2\eta \Gamma_{\text{ba}}}} = \sqrt{\left|\boldsymbol{\Sigma}_{c}^{-}\right|}\:, \\& \Sigma^{+}_{c,12} \approx 0.
\end{split}
\label{eq: common mode covariance approx}
\end{equation}
Substituting these expressions into Eq.~\eqref{eq: sigma negativity} and Eq.~\eqref{eq: symplectic eigenvalue}, we obtain
\begin{equation}    \sigma\approx\sqrt{\left|\boldsymbol{\Sigma}_{c}^{+}\right|}\text{Tr}\left|\boldsymbol{\Sigma}_{c}^{-}\right|=\sqrt{\left|\boldsymbol{\Sigma}_{c}^{-}\right|}\left(\lambda_{1}+\lambda_{2}\right)\:,
\label{eq: approximate sigma negativity}
\end{equation}
and the minimal symplectic eigenvalue
\begin{equation}
\tilde{\nu}=\sqrt{\sqrt{\left|\boldsymbol{\Sigma}_{c}^{+}\right|}\lambda_{2}}\:,
\label{eq: approximate symplectic eigenvalue}
\end{equation}
where $\lambda_{1} > \lambda_{2} > 0$ are the eigenvalues of $\boldsymbol{\Sigma}_{c}^{-}$. Now, generally $\boldsymbol{X}_\text{div}$ and $\boldsymbol{X}_\text{dec}$ are not orthogonal, and their period-averaged scalar product is given by
\begin{equation}
    \left< \boldsymbol{X}_\text{div}^{T} \boldsymbol{X}_\text{dec} \right> = \frac{16 \Omega_0^{2} + \Omega_{c}^{2}}{32 \Omega_{0}^{2}} \cos \left( 2 \phi \right).
\label{eq: averaged modes projection}
\end{equation}
However, for the case of $\Omega_{c} = 2\Omega_{-}$, the relative phase is $\pi/4$, and we can therefore approximate $\boldsymbol{X}_\text{div}$ and $\boldsymbol{X}_\text{dec}$ as orthogonal, which in turn, given the form of $\Sigma^{-}_{c}(t)$ (See Eq.~\eqref{eq: Conditional approximation}) implies that 
\begin{multline}
    \lambda_{2} \approx \alpha_{\text{dec}} \text{Tr} \left[ \Sigma_{\text{dec}}\left(t\right) \right] \\=\frac{\Gamma_{\text{tot}}}{8g_{1}\Omega_{-}}\left[\Omega_{0}+4g_{0}\sin^{2}\left(\Omega_{-}t-\frac{\pi}{4}\right)\right]\:,
\label{eq: lambda2}
\end{multline}
and thus 
\begin{equation}
\begin{split}
        \mathcal{E}_{n} & =-\frac{1}{2}\ln\left(\sqrt{\frac{\Gamma_{\text{tot}}}{2\eta\Gamma_{\text{ba}}}}\right)  \\&- \frac{1}{2}\ln\left[\frac{\Gamma_{\text{tot}}}{2g_{1}}\frac{\Omega_{0}}{\Omega_{-}}\left[1+\frac{4g_{0}}{\Omega_{0}}\sin^{2}\left(\Omega_{-}t-\frac{\pi}{4}\right)\right]\right].
\end{split}
\label{eq: approximate logarithmic negativity}
\end{equation}
Equation~\eqref{eq: approximate logarithmic negativity} shows that the logarithmic negativity is composed of two terms, corresponding to the two diverging and decaying modes that form the conditional covariance matrix. The first, static term arises from the common mode, and the second time-dependent term originates from the differential mode. For comparison, logarithmic negativity in the non-modulated, attractive interacting case is approximated as \cite{winkler2024stationary}
\begin{equation}
    \mathcal{E}_{n}^{\text{static}}=-\frac{1}{2}\ln\left(\sqrt{\frac{\Gamma_{\text{tot}}}{2\eta\Gamma_{\text{ba}}}}\right)-\frac{1}{2}\ln\left(2\frac{\Omega_{0}}{\Omega_{-}}\sqrt{\frac{\Gamma_{\text{tot}}}{2\eta\Gamma_{\text{ba}}}}\right)\:. 
\label{eq: static logarithmic negativity}
\end{equation}
By comparing Eq.~\eqref{eq: approximate logarithmic negativity} and Eq.~\eqref{eq: static logarithmic negativity}, one finds that the modulation enhances the entanglement by replacing the need for higher detection efficiency, $\eta$, with the parametric gain, characterized by $g_{1}$. Figure \ref{fig: g1-eta logneg} shows the logarithmic negativity as a function of the efficiency $\eta$ and the interaction modulation strength $g_{1}$ for both the numerical results and the analytical predictions of Eq.~\eqref{eq: approximate logarithmic negativity}. A good agreement can be seen between the two.

\begin{figure}[b]
\centering\includegraphics[width=0.49\textwidth]{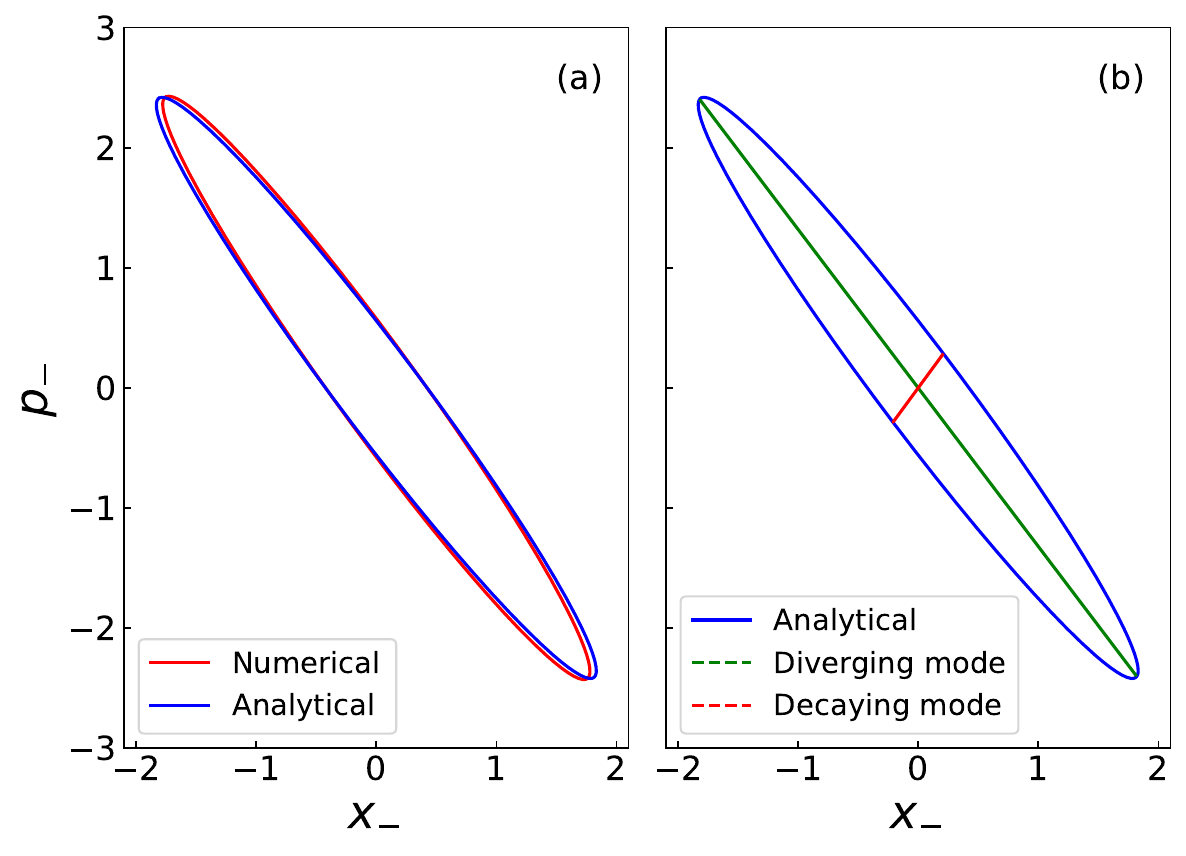}
\caption{(a)  Comparison of the ellipse-shaped, standard deviation of the differential mode's Wigner distribution between the numerical and analytical solutions. (b) Decomposition of the ellipse axes into the diverging and decaying modes. Calculated for the attractive case at the end of one modulation period for a frequency of $\Omega_{c} = 2\Omega_{-}$,  Parameters match Fig.~\ref{fig: g1-eta logneg}, except for $g_{1}/g_{0}=0.25$ and $\eta=0.5$.}
\label{fig: noise ellipse}
\end{figure}

Another aspect of the approximation is illustrated in Fig.~\ref{fig: noise ellipse}(a-b), which presents the characteristic, ellipse-shaped, standard deviation of the differential mode's Wigner distribution. The area of the ellipse is given by the determinant of the covariance matrix (Eq.~\eqref{eq: ellipse area}), and its axes are determined by its eigenvectors \cite{Serafini2017quantum}. Fig.~\ref{fig: noise ellipse}(a) shows the ellipse for both numerical and analytical models, while Fig.~\ref{fig: noise ellipse}(b) shows how the ellipse can be decomposed into the two approximately orthogonal diverging and decaying modes of Eq.~\eqref{eq: Conditional approximation}. The two axes are not fully orthogonal, since the two modes are orthogonal only on average (see Eq.~\eqref{eq: averaged modes projection}). Entanglement is achieved when one axis is squeezed below the appropriate threshold, essentially creating two-mode squeezing of the differential mode,  which manifests in a positive logarithmic negativity. For this reason, the logarithmic negativity depends only on the smaller eigenvalue $\lambda_{2}$, as shown in equation~\eqref{eq: approximate symplectic eigenvalue}.

We now proceed to the calculation of the unconditional entanglement.
Unlike the nonlinearity in Eq.~\eqref{eq: covariance_EOM}, the nonlinearity governing the dynamics of $\Omega(t)$ in Eq.~\eqref{eq: gain matrix eom} is typically much stronger than its linear contribution. Indeed, one usually finds $1/q > \Omega_{0}$, and in some cases even $1/q \gg \Omega_{0}$. As a result, one can neglect the time modulation of \eqref{eq: gain matrix eom}, and assume that the dynamics reach a steady state, as in \cite{winkler2024stationary}, and analytically solve the resulting algebraic Riccati equation
\begin{equation}
\begin{split}
\boldsymbol{A}\Omega + \Omega\boldsymbol{A}^{T} + \boldsymbol{P} - 
    \Omega \frac{\boldsymbol{B}\boldsymbol{B}^{T}}{q} \Omega=0\:,
\end{split}
\label{eq: approximated optimal feedback matrix EOM }
\end{equation}
where $\boldsymbol{A}$ is the static version of~\eqref{eq: Drift matrix}
\begin{equation}
    \boldsymbol{A} = \begin{pmatrix}0 & \Omega_{0}\\
    -\Omega_{0}-4g_{0}& 0 \end{pmatrix}. 
\label{eq: const A}
\end{equation}
Equation~\eqref{eq: excess noise EOM}, which describes the dynamics of the excess noise matrix, $\Xi (t)$, is linear and nonhomogeneous. Its solution can therefore be expressed as the sum of the homogeneous, Mathieu-like solutions, and a particular solution of the full nonhomogeneous equation. From the explicit form of the linear term in Eq.~\eqref{eq: excess noise EOM}, 
\begin{equation}
    \boldsymbol{A}\left(t\right)-\boldsymbol{B}\boldsymbol{K} = \begin{pmatrix}0 & \Omega_{0}\\
-\Omega_{0}-\frac{\Omega_{11}^{-}}{q}-4g\left(t\right) & -\gamma - \frac{\Omega_{22}^{-}}{q}
\end{pmatrix}\:,
\label{eq: A-BK}
\end{equation}
we see that the gain matrix contributes a friction term, represented by $\Omega_{22}/q$, alongside a frequency shift, characterized by $-\Omega_{11}^{-}/q$. Since  $1/q > \Omega_{0}$, the shift is sufficiently strong to move the system far from the main resonance window of twice the natural frequency. Moreover, according to the general properties of the parametric resonance~\cite{Kovacic2018Mathieu}, if the friction rate is of the order of the natural frequency or larger, the system will decay to zero regardless of the modulation. Therefore, we can neglect the homogeneous solutions and the modulation and consider inhomogeneous solution alone
\begin{multline}
d\Xi\left(t\right) =\left(\boldsymbol{A}-\boldsymbol{B}\boldsymbol{K}\right)\Xi\left(t\right)dt+\Xi\left(t\right)\left(\boldsymbol{A}-\boldsymbol{B}\boldsymbol{K}\right)^{T}dt \\+\Sigma^{c}\left(t\right)\boldsymbol{C}^{T}\boldsymbol{W}^{-1}\boldsymbol{C}\Sigma^{c}\left(t\right)dt\:,
\label{eq: approximated excess noise}
\end{multline}
 which is formally given by 
\begin{equation}
    \Xi(t) = e^{\left( \boldsymbol{A}-\boldsymbol{B}\boldsymbol{K} \right) t} \boldsymbol{R}(t)  e^{\left( \boldsymbol{A}-\boldsymbol{B}\boldsymbol{K} \right)^{T} t}, 
\label{eq: Xi approxaimted}
\end{equation}
with 
\begin{equation}
    \begin{split}
        \boldsymbol{R}(t) = &\int_{t_0}^t e^{-\left( \boldsymbol{A}-\boldsymbol{B}\boldsymbol{K} \right) t'}  \Sigma^{c}\left(t' \right)\boldsymbol{C}^{T}\boldsymbol{W}^{-1}\boldsymbol{C}\\&\times\Sigma^{c}\left(t
    '\right) e^{-\left( \boldsymbol{A}-\boldsymbol{B}\boldsymbol{K} \right)^{T} t'} dt'  + \boldsymbol{R} (t_0).
    \end{split}
\end{equation}
In another words, this means, that the enhancement of the unconditional negativity at the parametric resonance does not {only} happen directly due to the parametric modulation, but {also} because of the resonantly enhanced nonlinear term $\propto \Sigma^{c}\boldsymbol{C}^{T}\boldsymbol{W}^{-1}\boldsymbol{C}\Sigma^{c}$. This can explain the higher threshold for unconditional entanglement in Fig.~\ref{fig: g1-eta logneg}(b).

\begin{figure}
\centering\includegraphics[width=0.49\textwidth]{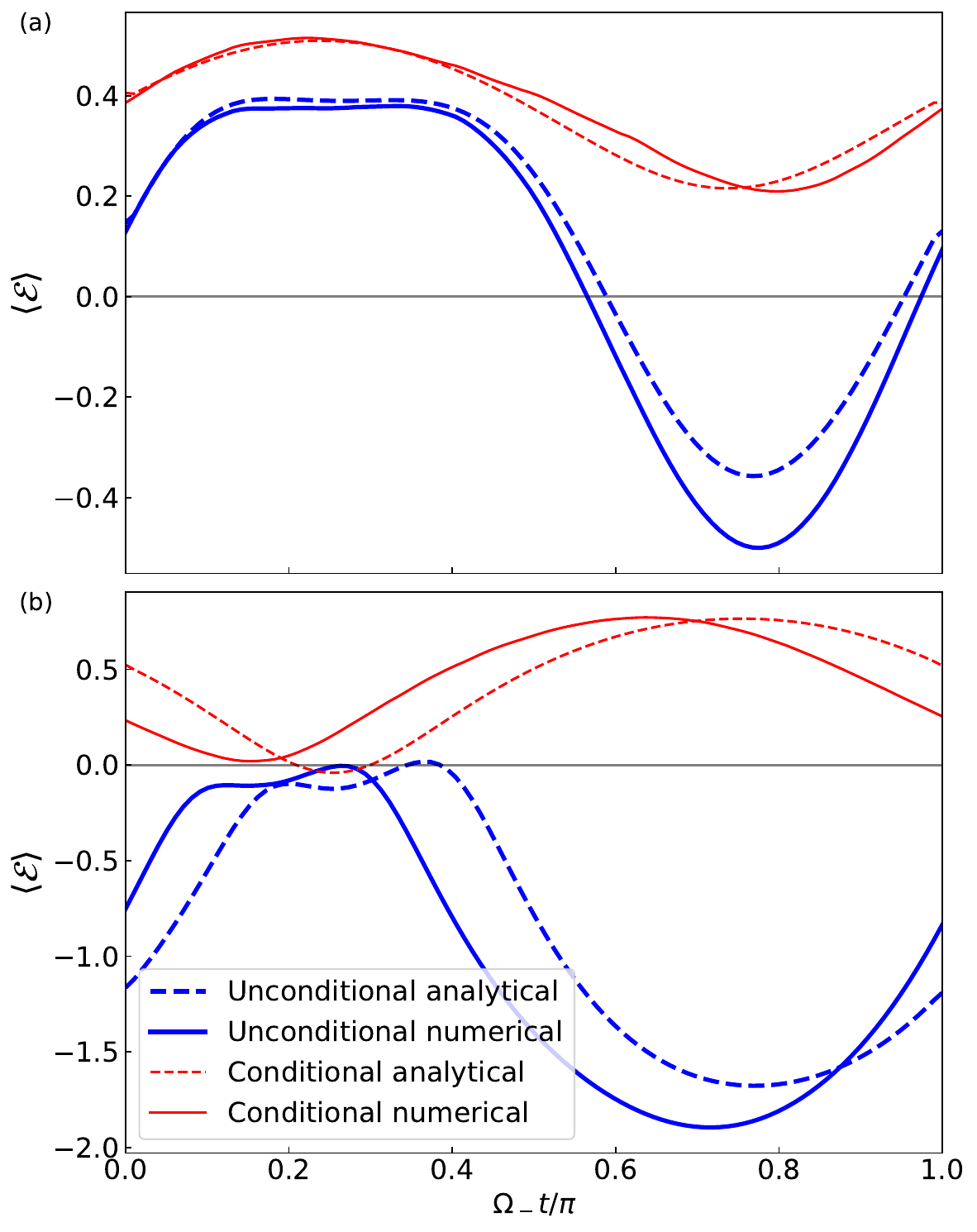}
\caption{Time evolution of the logarithmic negativity over one driving period for attractive interaction with  $g_0/\Omega_0 = 0.2$ (a), and for repulsive interaction with $g_0/\Omega_0 = -0.2$ (b). Both approximate (analytical) and numerical solutions are shown for the conditional and unconditional cases. The entanglement threshold is indicated by the black line at zero. System parameters are identical to those of Fig. \ref{fig: g1-eta logneg}, where we also set $g_{1}/|g_{0}| = 0.25$ and $\eta = 0.5$.} 
\label{fig: temporal log neg}
\end{figure}

Figure~\ref {fig: temporal log neg} presents a comparison of the analytical model against the numerical results for resonance frequency modulation of $\Omega_{c} = 2\Omega_{-}$ and modulation strength of $g_{1} / g_{0}=|0.2|$. Figure~\ref{fig: temporal log neg}(a) describes the attractive case, and Fig.~\ref{fig: temporal log neg}(b) describes the repulsive case. While a relatively good agreement is observed for the attractive case, the repulsive graphs show a larger phase deviation between analytics and numerics. At the same time, it can be seen that stroboscopic unconditional entanglement is achieved for the attractive interaction, contrary to the repulsive one. Both results can be understood from the Mathieu equation \eqref{eq: Mathieu_special}, particularly from the modulation depth $h=8\Omega g_{1} / \Omega_{-}^{2}$. The $\Omega_{-}^{2}$ factor in the denominator of $h$ is larger than $\Omega_{0}$ in the attractive case, and smaller in the repulsive. As a result, the amplitude of $h$ will decrease for the former and increase for the latter. For the repulsive parameters used in Fig.~\ref{fig: temporal log neg}(b), the modulation amplitude does not satisfy the condition $h \ll 1$, and thus the system will both deviate from the analytical model. At the same time, the large modulation amplitude will also cause the system to heat up and become less entangled.

\begin{figure}
\centering\includegraphics[width=0.49\textwidth]{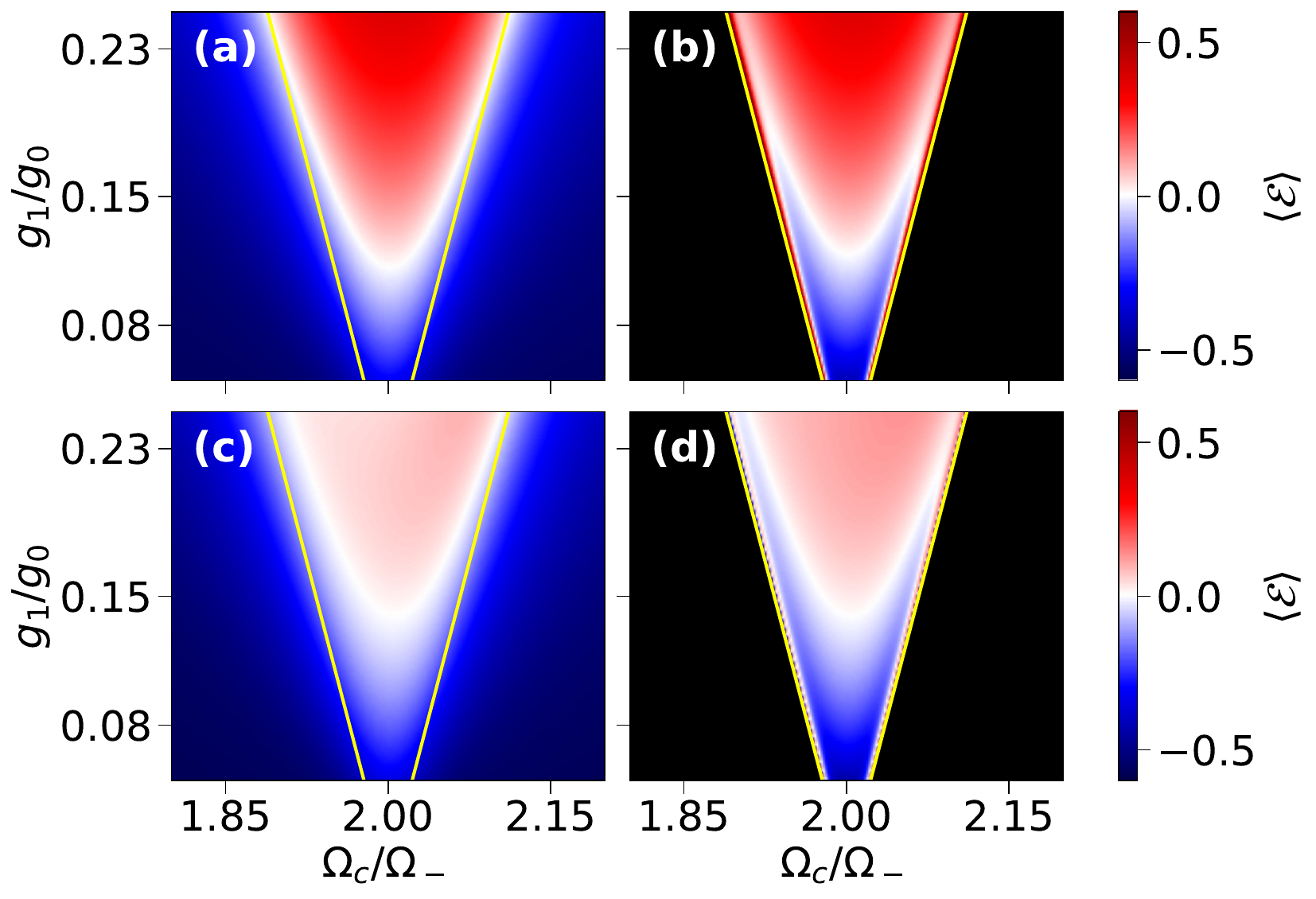}
\caption{Period-averaged logarithmic negativity as a function of the driving modulation frequency $\Omega_c$ and driving amplitude $g_1$. Panels (a) and (b) present the numerical and analytical conditional logarithmic negativity, respectively; panels (c) and (d) show the corresponding unconditional predictions. Black regions in (b) and (d) mark parameter ranges where the model is not applicable. The dashed yellow line signifies the theory-predicted resonance range (Eq.~\eqref{eq:range}) in all of the plots. Results are calculated for attractive interactions with $g_{0}/\Omega_{0}=0.2$. All parameters are the same as in Fig.~\ref{fig: g1-eta logneg}, with the addition of $g_{1}/g_{0} = 0.25$ and $\eta = 0.5$.}
\label{fig: logneg average attractive}
\end{figure}

The above results are illustrated further in Fig.~\ref{fig: logneg average attractive}, which presents, for the attractive interaction case, color maps of the average conditional and unconditional logarithmic negativity, averaged over one oscillation period. The maps show the  dependence of the logarithmic negativity on the modulation amplitude $\Omega_{c}$ and strength $g_{1}$ for both numerical and analytical results. The agreement between the analytical model and the numerical results is evident. The enhancement of the negativity takes place in the region of the parametric resonance,  $-\frac{1}{2}h<\epsilon<\frac{1}{2}h$, or, explicitly,
\begin{equation}\label{eq:range}
\left|\Omega_{\rm c}- 2\Omega_- \right|<\frac{4\Omega_0g_1}{\Omega_-}\:, 
\end{equation}
indicated by the yellow lines in Fig.~\ref{fig: logneg average attractive}. The behavior of the system for repulsive interactions is discussed further in Appendix~ 
\ref {app: Repulsive interaction}.

\section{Summary}\label{sec:summary}
To summarize,  we have put forward a theory for the parametric resonance enhancement of  continuous-variable entanglement of the motional degrees of freedom of two optically trapped particles under optimal control. Our model accurately reproduces the numerically calculated time dependence of the covariance matrix and logarithmic negativity in a wide range of parameters. 
 While it is generally expected that parametric driving will lead to the squeezing of the differential mode of the two particles, it is not immediately clear how to obtain quantitative results for a measure of entanglement. Here, we derive explicit, closed-form analytical expressions for the time dependence of the conditional negativity and present an analytical approach for the unconditional negativity. 
 We have explicitly identified the contributions of the common and differential Floquet modes to the generation of entanglement.
In particular, our analytical results for resonant parametric driving indicate that the conditional entanglement is primarily determined by the competition between the parametric gain and the total decay, rather than by the measurement efficiency. This is in stark contrast with the stationary case without parametric driving~\cite{winkler2024stationary}.  Our results provide a rigorous quantitative test of the intuition based on the squeezing of the differential mode and also indicate when this simplified description breaks down.
For the attractive interaction, $g_0>0$, the semi-analytical results agree with numerics up to the relative modulation strength $g_1/g_0=0.25$. For repulsive interaction, $g_0<0$, the applicability range of the semi-analytical model is narrower, which is explained by the lower differential mode frequency.

We hope that these findings provide useful insights to further optimize the experimental setups and achieve entanglement in the lab. Our theory is rather general and can potentially be applied to entanglement processes in other parametrically modulated quantum systems under optimal control.

\acknowledgments 

This work has been inspired by multiple conversations with Klemens Winkler and Anton Zasedatelev. We also thank Roni Ben Maimon, Nikita Leppenen, Oz Matoki, Eli Meril, Ephraim Shahmoon, and Yoav Shimshi for useful discussions. We acknowledge financial support by the Israel Science Foundation Quantum Program Grant No. 2197/24.

\appendix
\section{Eigenmodes  of the Mathieu equation}\label{app: modes derivation} 
In this Appendix, we recall the solutions for the decaying and diverging eigenmodes of the Mathieu equation  
\begin{equation}
\ddot{x}=-\omega^{2}\left\{1+h\cos\left[\left(2+\epsilon\right)\omega t\right]\right\}x\:,
\label{Aeq: Mathieu equation}
\end{equation}
that are used in the main text. The derivation is mostly based on Ref.~\cite{Kovacic2018Mathieu}. Assuming a weak and almost resonant modulation ($h,\epsilon \ll 1$), we apply the two variable expansion method \cite{Rand1987}, setting the fast time $\xi = (2 + \epsilon) \omega t$ and the slow time $\eta = h t $, such that 
\begin{equation}
    \frac{d^{2}}{dt^{2}} = \left(2+\epsilon\right)^{2}\omega^{2}\frac{\partial^{2}}{\partial\xi^{2}}+2\left(2+\epsilon\right)\omega h\frac{\partial^{2}}{\partial\xi\partial\eta}+\epsilon^{2}\frac{\partial^{2}}{\partial\eta^{2}}.
\label{Aeq: time derivative two variable expansion}
\end{equation}
We also expand $x$ in powers of $h$ as 
\begin{equation}
    x\left( \xi, \eta \right) = x_{0} \left( \xi, \eta \right) + h x_{1} \left( \xi, \eta \right) + ...
\label{aeq: x h expansion} 
\end{equation}
and take only the zeroth and first orders in $h,\epsilon \ll 1$ into account. Substituting~\eqref{aeq: x h expansion} into~\eqref{Aeq: Mathieu equation} gives to zeroth order in $h, \epsilon$
\begin{equation}
\begin{split}
    & 4\omega^{2}\frac{\partial^{2}x_{0}}{\partial\xi^{2}}=-\omega^{2}x_{0} \:,\\& \Rightarrow x_{0}=A\left(\eta\right)\cos\left(\frac{\xi}{2}\right)+B\left(\eta\right)\sin\left(\frac{\xi}{2}\right).
\end{split}
\label{Aeq: zeroth order}
\end{equation}
To find $A(\eta)$ and $B(\eta)$, we consider the first-order terms
\begin{equation}
     \frac{\partial^{2}x_{1}}{\partial\xi^{2}}+\frac{x_{1}}{4}=-\frac{\epsilon}{h}\frac{\partial^{2}x_{0}}{\partial\xi^{2}}-\frac{1}{\omega}\frac{\partial^{2}x_{0}}{\partial\xi\partial\eta}-\frac{1}{4}\cos\left(\xi\right)x_{0}.
\label{Aeq: first order EOM}
\end{equation}
Plugging~\eqref{Aeq: zeroth order} into the above results in
\begin{equation}
\begin{split}
    & \frac{\partial^{2}x_{1}}{\partial\xi^{2}}+\frac{x_{1}}{4}	=-\frac{\epsilon}{4h}\left(A\cos\frac{\xi}{2}+B\sin\frac{\xi}{2}\right)
	\\& +\frac{1}{2\omega}\left(\frac{dA}{d\eta} \sin\frac{\xi}{2}-\frac{dB}{d\eta}\cos\frac{\xi}{2}\right)
	\\& -\frac{1}{8}\left(A\cos\frac{\xi}{2} +A\cos\frac{3\xi}{2}+B\sin\frac{3\xi}{2}-B\sin\frac{\xi}{2}\right).
\end{split}
\label{Aeq: first order}
\end{equation}
To ensure the self-consistency of the perturbation, the condition $hx_{1}\ll x_{0}$ must hold. consequentially, the resonance terms, oscillating at frequency $\omega$, need to be removed by satisfying the slow time equations
\begin{equation}
    \frac{dA}{d\eta} = -\omega\left(\frac{1}{4}-\frac{\epsilon}{2 h}\right)B, \qquad \frac{dB}{d\eta} = -\omega\left(\frac{1}{4}+\frac{\epsilon}{2 h}\right)A.
\label{Aeq: slow time ODE}
\end{equation}
This results in solutions of the form
\begin{equation}
\begin{split}
       & A(t) = c_{1} e^{\frac{\mu t}{2}} + c_{2} e^{\frac{-\mu t}{2}}\:, \\& B(t) = \sqrt{\frac{2\epsilon-h}{2\epsilon+h}} \left( c_{1} e^{\frac{\mu t}{2}} - c_{2} e^{\frac{-\mu t}{2}}  \right),
\end{split}
\label{Aeq: slow time A B sol.}
\end{equation}  
with the Mathieu exponent $\mu=\omega\sqrt{\frac{h^{2}}{4}-\epsilon^{2}}$, and $c_{1,2}$ as the initial conditions coefficients. Due to the structure of the $\mu$, exponential growth and decay will appear for real $\mu$, limiting the resonance window to the frequency detuning range $-h/2 < \epsilon <h/2$. 
Next, we substitute~Eq.~\eqref{Aeq: slow time A B sol.} into~\eqref{Aeq: zeroth order}, and recover Eq.~\eqref{eq: Mathieu solutions} for the diverging and decaying modes from the main text:
\begin{equation}
\begin{split}
        & e^{\frac{\mu t}{2}} \boldsymbol{X}_{\text{div}} (t)= e^{\frac{\mu t}{2}} \begin{pmatrix}\cos\left(\frac{\Omega_{c}}{2}t+\phi\right)\\-\frac{\Omega_{c}}{2 \Omega_{0}}\sin\left(\frac{\Omega_{c}}{2}t+\phi\right)\end{pmatrix}\:,
        \\& e^{-\frac{\mu t}{2}} \boldsymbol{X}_{\text{dec}} (t) = e^{-\frac{\mu t}{2}}\begin{pmatrix}\cos\left(\frac{\Omega_{c}}{2}t-\phi\right)\\ -\frac{\Omega_{c}}{2 \Omega_{0}}\sin\left(\frac{\Omega_{c}}{2}t-\phi\right)\end{pmatrix}.
\end{split}
\label{Aeq: Mathieu solutions1}
\end{equation}
Here, 
$\bm X\equiv (x,p)^T$ with $p=(1/\omega)dx/dt$ being the conjugate momentum,
$\tan \phi = \sqrt{\frac{h - 2\epsilon}{h + 2\epsilon}}$, 
 $\omega=\Omega_{-}$, $h=8\Omega_{0} g_{1} / \Omega_{-}^{2}$, and $\Omega_{c} = \left(2 +\epsilon\right) \Omega_{-} $.

\begin{figure}[t!]
\centering\includegraphics[width=0.48\textwidth]{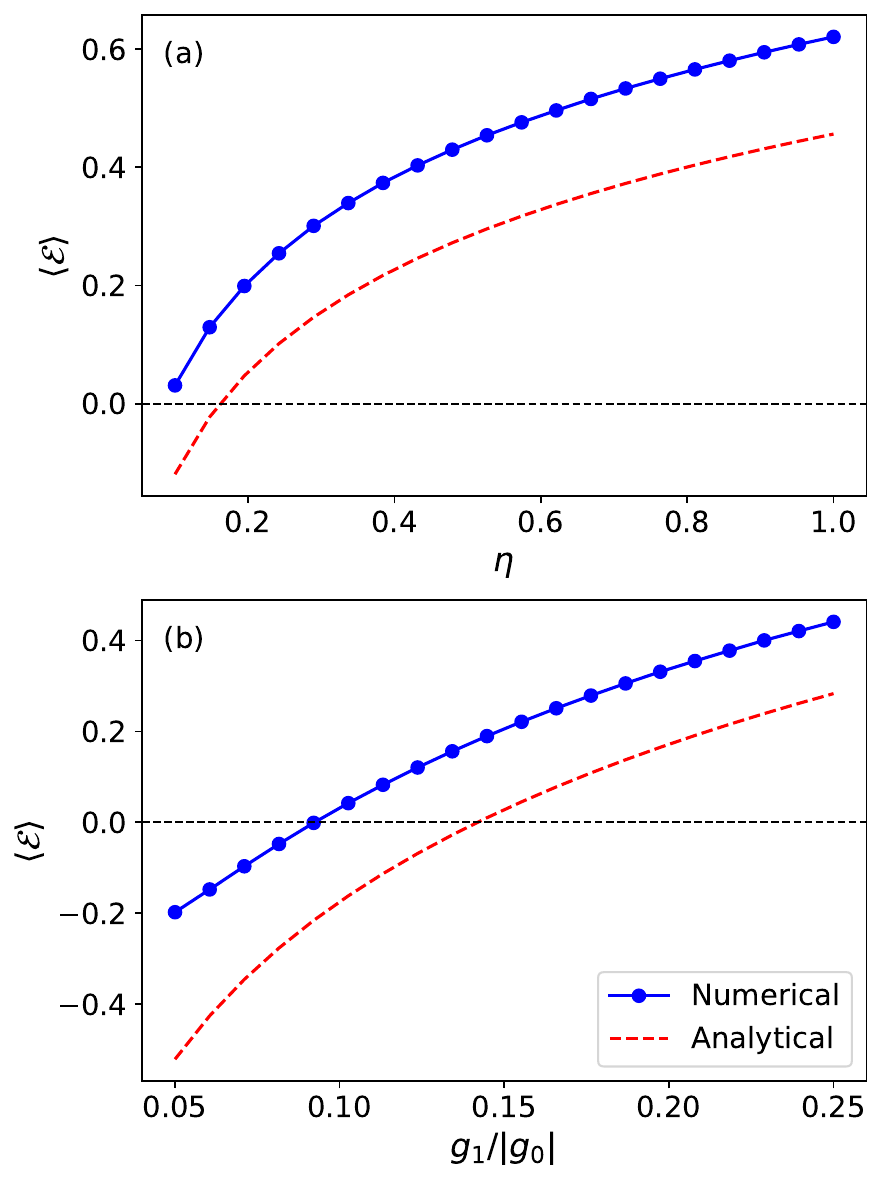}
\caption{Similar to Fig.~\ref{fig: g1-eta logneg}, but for 
repulsive interaction.  Logarithmic negativity for the modulation frequency  $\Omega_{c}=2 \Omega_{-}$ as a function of the efficiency $\eta$ (a) and the modulation depth $g_{1}$ (b) averaged over one period. Solid line with markers denotes the numerical results, and the dashed line describes the value of the analytical expression of the negativity Eq.~\eqref{eq: approximate logarithmic negativity}. The black dashed line indicates the entanglement threshold. Apart from $g_{0}$ being negative, the parameters are the same as in Fig. \ref{fig: g1-eta logneg}.}
\label{fig_app: g1-eta logneg}
\end{figure}

\begin{figure}[t!]
\centering\includegraphics[width=0.48\textwidth]{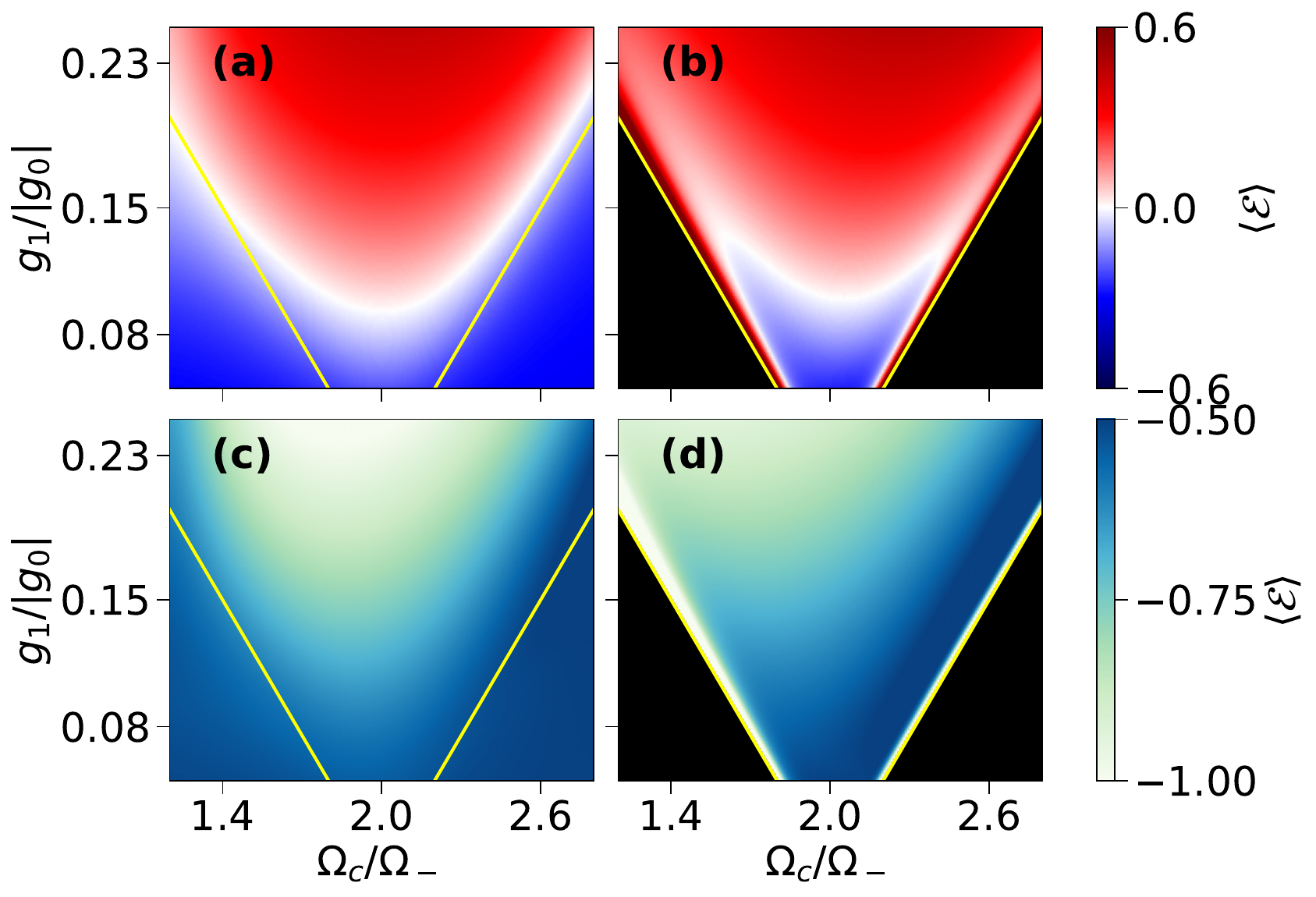}
\caption{Similar to Fig.~\ref{fig: logneg average attractive}, but for 
repulsive interaction. Color maps present period-averaged logarithmic negativity as a function of the driving modulation frequency $\Omega_c$ and driving amplitude $g_1$. Panels (a) and (b) present the numerical and analytical conditional logarithmic negativity, respectively; panels (c) and (d) show the corresponding unconditional results. Black regions in (b) and (d) mark parameter ranges where the model is not applicable. The dashed yellow line indicates the theory-predicted resonance range Eq.~\eqref{eq:range} in all of the plots. Different color maps are used for each line due to the different logarithmic negativity ranges. Results are calculated for attractive interactions with $g_{0}/\Omega_{0}=-0.2$. All parameters are the same as in Fig.~\ref{fig: g1-eta logneg}, with the addition of $|g_{1}/g_{0}| = 0.25$ and $\eta = 0.5$.}
\label{fig_app: logneg average repulsvie} 
\end{figure}

\section{Model performance --- repulsive interaction}\label{app: Repulsive interaction}

The performance of the analytical model is sensitive to the sign of $g_{0}$. As discussed in Sec.~\ref{sec:analytics}, a repulsive interaction $(g_{0} < 0)$ increases the parameter $h$, which in turn reduces both the accuracy of the approximation in Eq.~\eqref{eq: Mathieu solutions} and the logarithmic negativity itself due to heating. Nevertheless, our model still captures the qualitative behavior of conditional logarithmic negativity, as shown in Fig.~\ref{fig: temporal log neg}(b). The deviation from the numerical results is mostly manifested in a phase delay of the logarithmic negativity and noise ellipse curves, the mechanism of which requires further investigation. This phase delay decreases the approximated averaged orthogonality between the two modes [see Eq.~\eqref{eq: averaged modes projection}], thereby reducing the validity of the closed-form expression for the negativity Eq.~\eqref{eq: approximate logarithmic negativity}, as illustrated in Fig.~\ref{fig_app: g1-eta logneg}.  The latter shows a wider, relatively constant discrepancy between the analytical theory and the numerics. Despite this, the overall shapes of the graphs remain identical, indicating that the $\Gamma_{\text{ba}}/\eta$  scaling of the conditional negativity, observed for the attractive interaction in Fig.~\ref{fig: g1-eta logneg}, seems to be valid for the repulsive regime as well. Unlike the conditional case, the unconditional negativity exhibits not only a phase discrepancy, but also a more pronounced deviation as the modulation frequency detunes from the $2 \Omega_{-}$ resonance, as demonstrated in Fig.~\ref{fig_app: logneg average repulsvie}. A comparison between  Fig.~\ref{fig_app: logneg average repulsvie}(c) and Fig.~\ref{fig_app: logneg average repulsvie}(d) shows that the analytically calculated color map is spectrally shifted towards lower frequencies relative to the numerical graph. The cause of the shift is likely due to the dynamical phase sensitivity of excess noise, presented in the last term of Eq.~\eqref{eq: excess noise EOM}. We note that the asymmetry of Fig.~\ref{fig_app: logneg average repulsvie} is a feature present already in Mathieu equation theory \cite{Kovacic2018Mathieu}, and is likely not related to the nonlinearity or the analytical model.  

\FloatBarrier
\bibliography{entanglement}

\end{document}